\documentclass[aps,prl,twocolumn,superscriptaddress,floatfix]{revtex4-2}

\usepackage[utf8]{inputenc}
\usepackage{graphicx,xcolor}
\usepackage{amsmath}
\usepackage{etoolbox}

\begin{document}


\title{A phase diagram for light-induced superconductivity in $\kappa$-(ET)$_2$-X}


\author{M. Buzzi}
\email[]{michele.buzzi@mpsd.mpg.de}
\affiliation{Max Planck Institute for the Structure and Dynamics of Matter, 22761 Hamburg, Germany}

\author{D. Nicoletti}
\affiliation{Max Planck Institute for the Structure and Dynamics of Matter, 22761 Hamburg, Germany}

\author{S. Fava}
\affiliation{Max Planck Institute for the Structure and Dynamics of Matter, 22761 Hamburg, Germany}

\author{G. Jotzu}
\affiliation{Max Planck Institute for the Structure and Dynamics of Matter, 22761 Hamburg, Germany}

\author{K. Miyagawa}
\affiliation{Department of Applied Physics, University of Tokyo, 7-3-1 Hongo, Bunkyo-ku, Tokyo, Japan}

\author{K. Kanoda}
\affiliation{Department of Applied Physics, University of Tokyo, 7-3-1 Hongo, Bunkyo-ku, Tokyo, Japan}

\author{A. Henderson}
\affiliation{National High Magnetic Field Laboratory, 1800 E Paul Dirac Drive, Tallahassee, FL 31310, USA}

\author{T. Siegrist}
\affiliation{National High Magnetic Field Laboratory, 1800 E Paul Dirac Drive, Tallahassee, FL 31310, USA}

\author{J. A. Schlueter}
\affiliation{National High Magnetic Field Laboratory, 1800 E Paul Dirac Drive, Tallahassee, FL 31310, USA}
\affiliation{Division of Material Research, National Science Foundation, Alexandria, VA 22314, USA}

\author{M.-S. Nam}
\affiliation{Department of Physics, Clarendon Laboratory, University of Oxford, Oxford OX1 3PU, United Kingdom}

\author{A. Ardavan}
\affiliation{Department of Physics, Clarendon Laboratory, University of Oxford, Oxford OX1 3PU, United Kingdom}

\author{A. Cavalleri}
\email[]{andrea.cavalleri@mpsd.mpg.de}
\affiliation{Max Planck Institute for the Structure and Dynamics of Matter, 22761 Hamburg, Germany}
\affiliation{Department of Physics, Clarendon Laboratory, University of Oxford, Oxford OX1 3PU, United Kingdom}

\date{\today}

\begin{abstract}
Resonant optical excitation of certain molecular vibrations in $\kappa$-(BEDT-TTF)$_2$Cu[N(CN)$_2$]Br has been shown to induce transient superconducting-like optical properties at temperatures far above equilibrium T$_c$. Here, we report experiments across the bandwidth-tuned phase diagram of this class of materials, and study the Mott-insulator $\kappa$-(BEDT-TTF)$_2$Cu[N(CN)$_2$]Cl and the metallic compound $\kappa$-(BEDT-TTF)$_2$Cu(NCS)$_2$. We find non-equilibrium photo-induced superconductivity only in $\kappa$-(BEDT-TTF)$_2$Cu[N(CN)$_2$]Br, indicating that the proximity to the Mott insulating phase and possibly the presence of pre-existing superconducting fluctuations are pre-requisites for this effect.
\end{abstract}


\maketitle

\begin{figure}[htbp]
	\includegraphics[width=0.95\columnwidth]{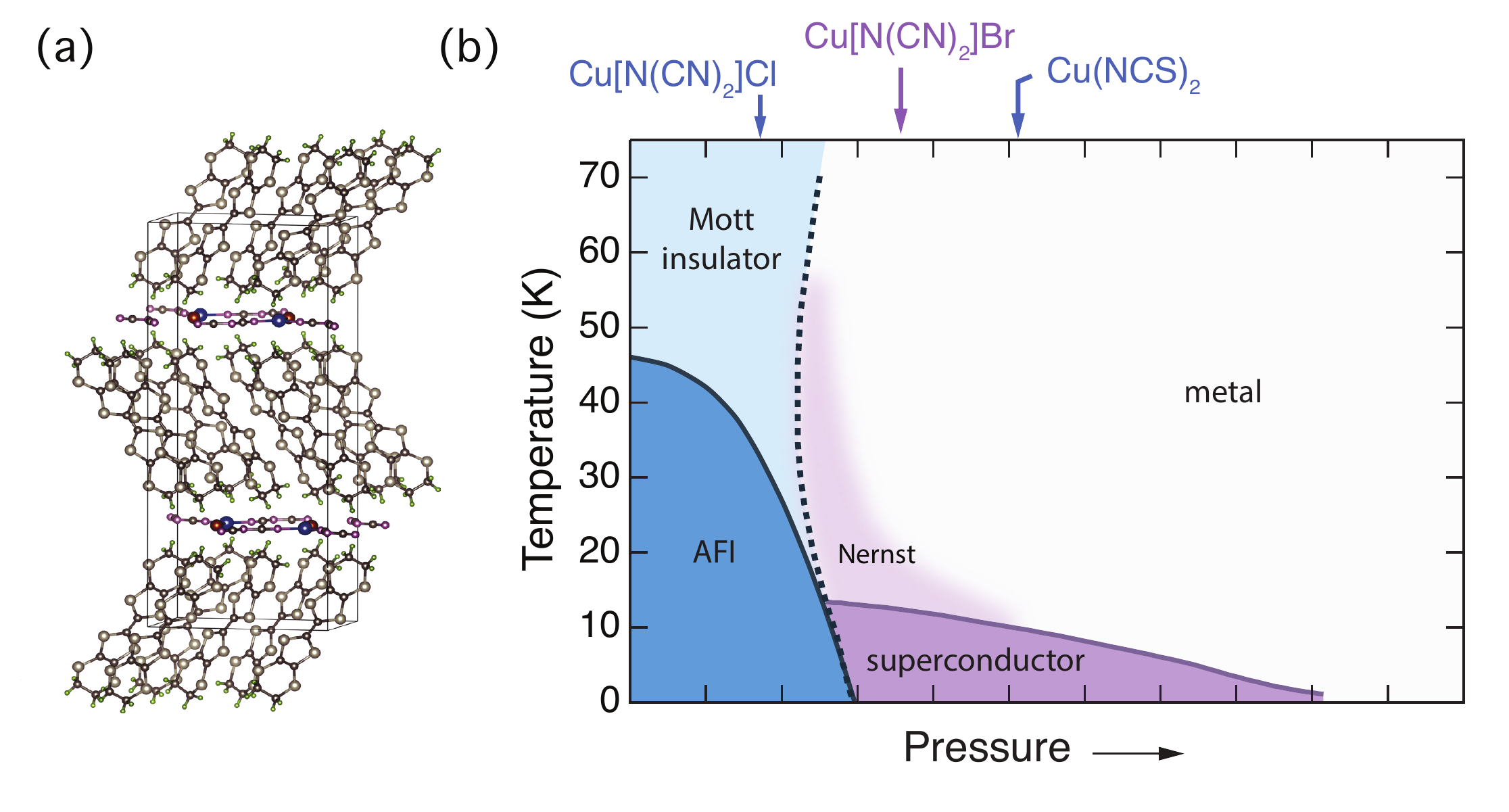}
    \caption{(a) Crystal structure of the $\kappa$-(ET)$_2$-X organic salt. (b) Temperature - effective pressure phase diagram of $\kappa$-(ET)$_2$-X. The three compounds studied in this work are highlighted along the horizontal axis. In proximity with the Mott insulating phase (light purple shading) measurements of Nernst effect reveal the presence of superconducting fluctuations above T$_c$ \cite{Nam2007,Nam2013}.\label{fig:Fig1}}
\end{figure}

Synthetic metals of the $\kappa$-(BEDT-TTF)$_2$X family exhibit high temperature unconventional superconductivity \cite{Ishiguro1998,Jerome1991,Lang2003} and bear some parallels with the physics of high-T$_c$ cuprates. In these materials, BEDT-TTF (bisethylenedithio-tetrathiafulvalene, henceforth abbreviated as ET) molecules are paired in dimers and stacked in layers to form a triangular lattice (Fig. \ref{fig:Fig1}(a)). Each one of the ET dimers donates an electron to the anion molecules X, which act as a spacer layer, resulting in half-filled conduction bands. The $\kappa$-(ET)$_2$X phase diagram (Fig. \ref{fig:Fig1}(b)) can be explored either by hydrostatic pressure \cite{Lefebvre2000,Kagawa2005} or by anion substitution \cite{Kanoda1997}. The compound with X=Cu[N(CN)$_2$]Cl ($\kappa$-Cl) is a Mott insulator, the one with X=Cu[N(CN)$_2$]Br ($\kappa$-Br), in close proximity with the Mott boundary, is a superconductor with the highest T$_c$ $\approx$ 12K  for this family of materials, and the one with X=Cu(NCS)$_2$ ($\kappa$-NCS) is also a superconductor with slightly lower transition temperature (T$_c$ $\approx$ 10K). Because of the layered structure, the normal state optical properties resemble those of an insulator across the ET layers \cite{McGuire2001}, whereas parallel to the planes the optical conductivity is insulating for the Mott phase of $\kappa$-Cl and metallic for $\kappa$-Br and $\kappa$-NCS. Notably, while for $\kappa$-NCS a vortex-Nernst effect is present only in the superconducting state, close to the Mott insulating phase ($\kappa$-Br) a vortex-like effect persists up to temperatures far above the superconducting T$_c$, which is suggestive of fluctuating superconductivity in the normal state \cite{Nam2007,Nam2013}.

\begin{figure*}[htbp]
	\includegraphics[width=0.75\textwidth]{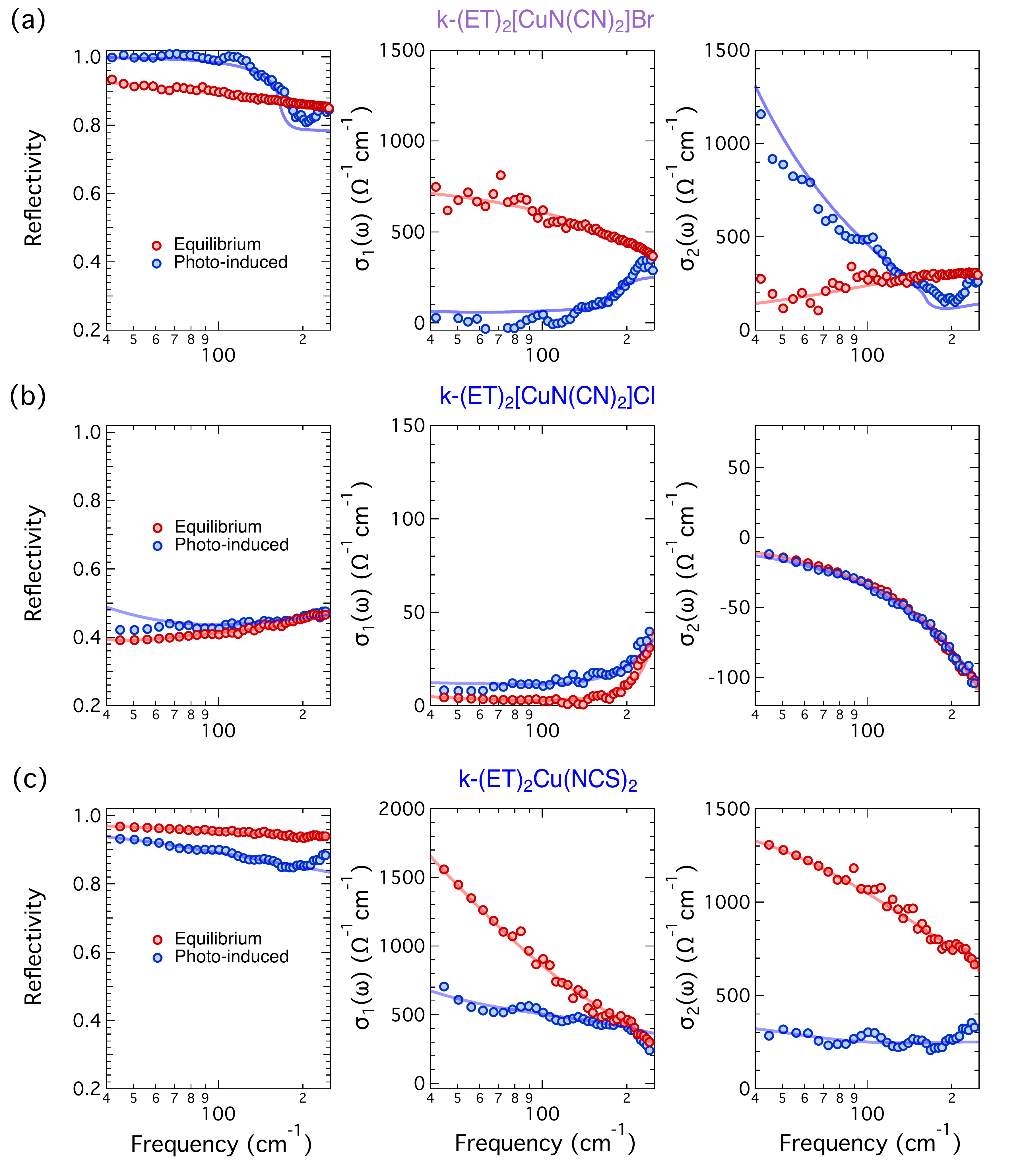}
    \caption{(a) In-plane reflectivity, real and imaginary part of the optical conductivity measured in $\kappa$-Br at equilibrium (red circles) and at $\tau \simeq$ 1 ps time delay after excitation (blue circles), at T = 30K. Solid lines are fits to the data with a Drude-Lorentz model (red) or a Mattis-Bardeen model for superconductors (blue). These data are reproduced from \cite{Buzzi2020}. (b) Same quantities as in (a) measured in $\kappa$-Cl at equilibrium (red circles) and at $\tau \approx$ 1 ps after excitation (blue circles), at T = 20K. Solid lines are fits to the optical spectra, which were performed with a Drude-Lorentz model for both the equilibrium and transient response. The equilibrium data are reproduced from \cite{Faltermeier2007}. (c) Same quantities measured for the metallic compound $\kappa$-NCS. Here, the equilibrium response was determined on the same crystal via Fourier-transform infrared spectroscopy. All data have been taken upon vibrational excitation close to resonance with the $\nu_{27}$ C=C stretching mode with a pump fluence of $\approx$3 mJ/cm$^2$\label{fig:Fig2}}
\end{figure*}

Recent work has focused on dynamical driving of $\kappa$-Br with intense laser pulses in the mid-infrared \cite{Buzzi2020}. Figure \ref{fig:Fig2}(a) summarizes the results of this experiment. Single crystals of $\kappa$-Br were cooled to temperatures for which the equilibrium response was that of a metal (here we show data taken at 30K) and driven with optical pulses tuned close to resonance with a C=C stretching mode of the ET molecules. Their in-plane non-equilibrium optical properties were then probed with phase-sensitive THz time-domain spectroscopy, yielding a response reminiscent of that of a superconductor with a perfect (R $\approx$ 1) reflectivity, a gap in the real part of the optical conductivity $\sigma_1(\omega)$, and a $\approx 1/\omega$ divergence in its imaginary part  $\sigma_2(\omega)$. These superconducting-like optical features were observed for all temperatures $T \leq T^* \simeq$ 50K at which the equilibrium normal state of $\kappa$-Br is a highly coherent quasi-two-dimensional Fermi liquid \cite{Dressel2010, Dumm2009}.

These experiments follow a number of qualitatively similar set of observations made in cuprates \cite{Fausti2011,Hu2014,Nicoletti2014,Cremin2019} and in fullerides \cite{Mitrano2016, Cantaluppi2018, Budden2021}. Microscopic explanations for this class of phenomena have ranged from the transient quasi static lattice distortions induced by non-linear lattice vibrations \cite{Mankowsky2014}, manipulation of competing orders \cite{Foerst2014} and the effect of dynamical modulations of the Hamiltonian parameters \cite{Singla2015,Buzzi2020}. Recent work has highlighted the ability to cool, amplify or otherwise manipulate normal state superconducting fluctuations \cite{Denny2015,Hoegen2020,Michael2020}. It has been proposed \cite{Uemura2019} that a prerequisite for light-induced superconductivity should be the presence of a phase-incoherent bosonic fluid in the normal state, where superconducting fluctuations would already be present. Here we explore this correlation by studying the response of two additional $\kappa$-(ET)$_2$X compounds, the Mott-insulating $\kappa$-Cl, and the superconducting $\kappa$-NCS, neither one exhibiting an anomalous vortex-like Nernst effect in their normal state. We find that in both materials the same vibrational excitation that was used for $\kappa$-Br does not result in a superconducting-like state.

Single crystals of $\kappa$-Cl and $\kappa$-NCS with typical dimensions of 0.5×0.5×0.3 mm$^3$ were synthesized by electro-crystallization and mounted on cone shaped holders to expose a surface that contained both the out-of-plane and one of the in-plane crystallographic directions. The crystals were photoexcited using ultrashort mid-infrared pump pulses generated using an optical parametric amplifier (OPA) pumped with amplified femtosecond pulses from a Ti:Sa laser. These pump pulses were polarized along the out-of-plane crystallographic axis and tuned close to resonance with the $\nu_{27}$ C=C stretching mode of the ET molecules. Broadband THz probe pulses ($\approx$1.2 to 7 THz) were generated in a 200-$\mu$m thick GaP (110) crystal from the direct output of the Ti:Sa amplifier (800nm wavelength). These THz probe pulses, with polarization parallel to the ET layers, were then focused on the sample and detected by electro-optic sampling after reflection in a second 200-$\mu$m thick GaP (110) crystal, yielding the photo-induced changes in the low-frequency complex reflection coefficient $r(\omega)$ as a function of pump-probe time delay. In $\kappa$-Cl the penetration depth of the mid-infrared pump (5.6 $\mu$m) was shorter than that of THz probe (7-20 $\mu$m). This was taken into account by modelling the sample as a multi-layered photo-excited stack on top of an unperturbed bulk in order to obtain the optical response functions corresponding to an effective semi-infinite and homogeneously excited medium \cite{Cantaluppi2018}. This procedure was not needed for $\kappa$-NCS where the pump pulses penetrated further in the material than the probe pulses.

Figures 2(b,c) illustrate the main findings of this paper. We report spectra of the optical properties ($R(\omega)$, $\sigma_1(\omega)$, and $\sigma_2(\omega)$), measured at equilibrium (red filled symbols) and 1 ps after photo-excitation (blue filled symbols) for $\kappa$-Cl (Fig. \ref{fig:Fig2}(b)) and $\kappa$-NCS (Fig. \ref{fig:Fig2}(c)). These measurements were performed at a base temperature T = 20K following the same excitation protocol that was used for $\kappa$-Br (Fig. \ref{fig:Fig2}(a)). The in-plane equilibrium spectra reported for $\kappa$-Cl show a low, featureless reflectivity and a vanishingly small real part of the optical conductivity $\sigma_1(\omega)$ both indicative of the insulating nature of this compound. $\kappa$-NCS shows, instead, a very different response, with a high reflectivity (R $\geq$ 0.9) and a broad Drude absorption in the optical conductivity, indicative of a metallic ground state. After photoexcitation, $\kappa$-Cl displays a slight, mostly frequency independent, enhancement in both reflectivity and optical conductivity. A more pronounced effect is seen in $\kappa$-NCS where photoexcitation at this temperature induces a reduction of the reflectivity, as well as a suppression and broadening of both $\sigma_1(\omega)$ and $\sigma_2(\omega)$. No signatures of light-induced superconductivity could be observed at T = 20K for both $\kappa$-Cl and $\kappa$-NCS, and all optical spectra (at equilibrium and after photoexcitation) could be captured by a Drude-Lorentz model with a finite scattering rate (red and blue solid lines, respectively).

\begin{figure}[htbp]
	\includegraphics[width=0.95\columnwidth]{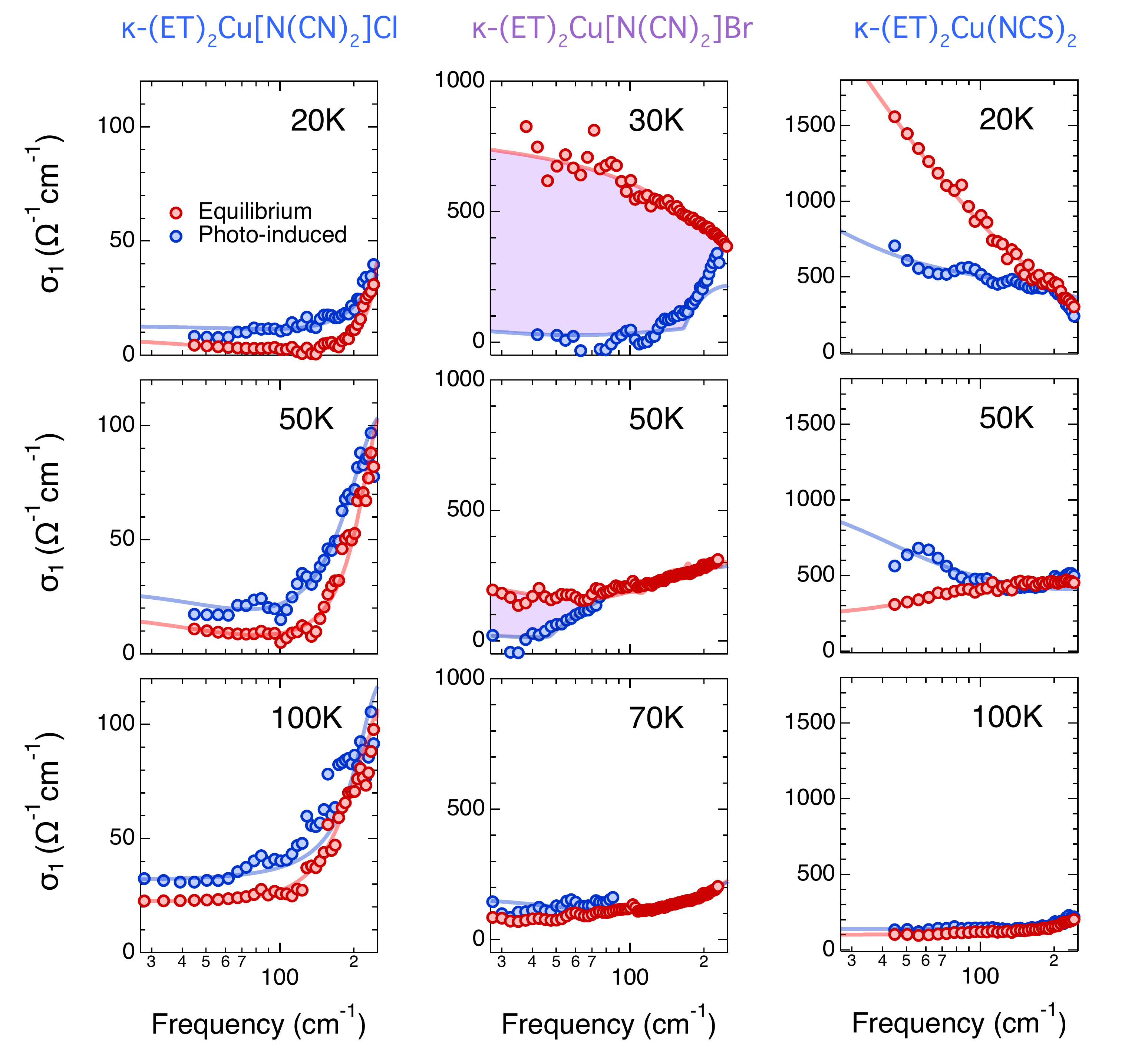}
    \caption{(left column) Real part of the optical conductivity measured in $\kappa$-Cl at equilibrium (red circles) and at $\tau \simeq$ 1ps after vibrational excitation (blue circles), at temperatures between 20K and 100K. The solid lines are fits to the optical spectra with a Drude-Lorentz model for both the equilibrium and out-of-equilibrium response. (center column) Same quantity as in the left column, measured in $\kappa$-Br at temperatures between 30K and 70K. Shaded areas indicate the lost spectral weight as a superconducting-like gap appears after optical excitation. Here, a Mattis-Bardeen model for superconductors was used for the out-of-equilibrium response at temperatures T $\leq$ 50 K. These data are reproduced from Ref. \cite{Buzzi2020}. (right column) Same quantities measured in $\kappa$-NCS at temperatures between 20K and 100K. The optical spectra were all modelled with a Drude-Lorentz fit (solid lines). All data were taken at a pump fluence of $\approx$3 mJ/cm$^2$\label{fig:Fig3}}
\end{figure}

A more complete view on the effect of photoexcitation in the $\kappa$-(ET)$_2$X compounds is offered by figure \ref{fig:Fig3} where we report the real part of the optical conductivity, $\sigma_1(\omega)$, measured at three different temperatures in $\kappa$-Cl, $\kappa$-Br, and $\kappa$-NCS under similar excitation conditions. For $\kappa$-Br, we observe that for all temperatures $T \leq T^* \approx$ 50K a clear superconducting-like gap opens at low frequencies, becoming progressively larger with decreasing temperature. This effect has been interpreted in terms of the onset of a photo-induced superconducting response \cite{Buzzi2020} and fitted with an extension of the Mattis-Bardeen model for superconductors. At T = 70K the response is qualitatively different: rather than the opening of a superconducting-like gap one observes an increase in $\sigma_1(\omega)$, indicative of enhanced metallicity.

In the case of insulating $\kappa$-Cl, we measure at all temperatures a slight increase of spectral weight, that one may attribute to photo-generation of free carriers. For metallic $\kappa$-NCS instead, at all T $\geq$ 50 K we observe an increase in $\sigma_1(\omega)$, similar to that found in $\kappa$-Br for T $\geq$ 70 K, thus indicative of enhanced metallicity. The photo-induced changes are different instead at T = 20K , where we observe a suppression in $\sigma_1(\omega)$ and a broadening of the Drude peak, possibly related to carrier heating. Hence, for both $\kappa$-Cl and $\kappa$-NCS the transient optical properties at all measured temperatures do not show any superconducting-like features and can be fully captured by the same Drude-Lorentz model used for the equilibrium optical spectra, with slightly varied plasma frequencies and scattering rates.

\begin{figure}[htbp]
	\includegraphics[width=0.95\columnwidth]{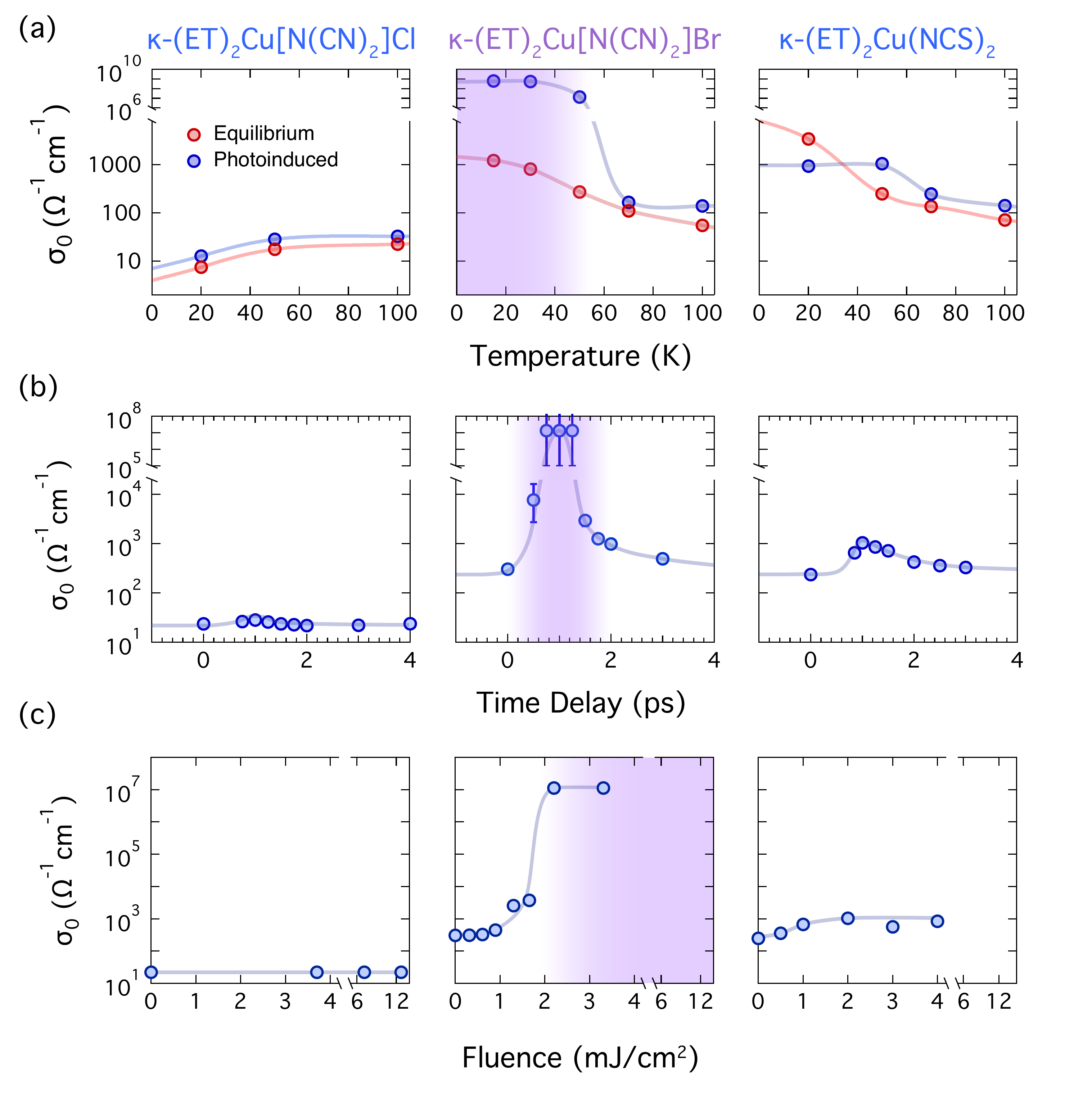}
    \caption{(a) Temperature dependence of $\sigma_0 = \sigma_1\rvert_{\omega \rightarrow 0}$ (see text) in transient (blue symbols) and equilibrium (red symbols) state for $\kappa$-Cl, $\kappa$-Br, and $\kappa$-NCS. (b) Time evolution of the same quantity as in (a), measured at a base temperature T = 50K. Data in (a,b) were taken upon  vibration excitation at a pump fluence of $\approx$3 mJ/cm$^2$. (c) Pump fluence dependence of $\sigma_0$ measured at $\tau \simeq$ 1 ps time delay after vibrational excitation in the same compounds. The data for $\kappa$-Br and $\kappa$-NCS were acquired at T = 50K while those for $\kappa$-Cl were taken at 20K. Solid lines are guides to the eye. Shaded areas indicate the regions of the parameter space where superconducting-like properties are found. The $\kappa$-Br data are reproduced from Ref. \cite{Buzzi2020}.\label{fig:Fig4}}
\end{figure}

By way of a summary, in figure \ref{fig:Fig4}(a) we report for all three samples the temperature dependence of the quantity $\sigma_0 = \lim_{\omega \rightarrow 0}\sigma_1(\omega)$, i.e. the extrapolated “zero-frequency” conductivity extracted from Drude-Lorentz fits to the transient (blue) and equilibrium (red) spectra (data in Figs. 2-3). Before photoexcitation (red), $\kappa$-Cl shows the typical temperature dependence expected for an insulator, where the low-frequency conductivity increases with increasing temperature due to thermally-activated carriers. In $\kappa$-Br and $\kappa$-NCS the behavior is instead opposite, as expected in a metal. Notably, whilst in $\kappa$-Cl and $\kappa$-NCS $\sigma_0$ remains finite at all temperatures, in $\kappa$-Br at $T \leq T^* \approx$ 50K photoexcitation causes $\sigma_0$ to diverge, compatible with the onset of dissipationless transport. In figure \ref{fig:Fig4}(b) we report the time dependence of the same quantity, $\sigma_0$, measured at T = 50K. In all three compounds, this changes promptly upon photoexcitation and relaxes over a few picoseconds, a time scale which is likely related to the lifetime of the driven vibrational mode. Finally, figure \ref{fig:Fig4}(c) shows the pump fluence dependent response: whilst in $\kappa$-Br, for all fluences F $\geq2\text{mJ/cm}^2$, $\sigma_0$ diverges to values compatible with a perfect conductivity, in $\kappa$-Cl and $\kappa$-NCS it always remains finite, with a significantly smoother dependence.
Our observations indicate that the presence of superconducting fluctuations in the normal state and proximity to a Mott insulating state correlate to the appearance of a photo-induced state with superconducting-like optical properties. These findings are also broadly compatible with the model we put forward previously \cite{Buzzi2020,Tindall2020} where a periodic modulation of the Hubbard interaction parameters yielded long range doublon correlations that may result in the generation of $\eta$-pairs and are expected to appear only below a certain value of the vertical hopping integral. One can speculate that this boundary may lay between $\kappa$-Br and $\kappa$-NCS. 
Another recent theoretical proposal \cite{Dai2021} discusses the appearance of superconducting-like optical properties after photo-excitation as a result of a non-equilibrium bosonic condensation of doubly occupied states, causing the simultaneous opening of a charge gap. Although within this scenario one would expect a superconducting-like response also in the insulating compound, which is not in direct agreement with our observation, this alternative interpretation highlights the importance of the vicinity to the Mott-insulating state. Note also that other measurements performed in K$_3$C$_{60}$ under pressure underscore the notion that the light induced superconducting-like response correlates with the electronic bandwidth of the material, becoming stronger as one approaches the Mott insulator \cite{Cantaluppi2018}.

\begin{acknowledgments}
The research leading to these results received funding from the European Research Council under the European Union’s Seventh Framework Programme (FP7/2007-2013)/ERC Grant Agreement No. 319286 (QMAC). We acknowledge support from the Deutsche Forschungsgemeinschaft (DFG, German Research Foundation) via the excellence cluster ‘The Hamburg Centre for Ultrafast Imaging’ (EXC 1074 – project ID 194651731) and the priority program SFB925 (project ID 170620586). J. A. Schlueter acknowledges support from the Independent Research/Development program while serving at the National Science Foundation. K. Miyagawa and K. Kanoda acknowledge support from the Japan Society for the Promotion of Science Grant No. 18H05225, 19H01846, 20K20894 and 20KK0060.
\end{acknowledgments}

\end{document}